\documentclass[twocolumn,reprint,aps,prb, floatfix,amsmath,amssymb,superscriptaddress]{revtex4-2}

\usepackage{graphicx}
\usepackage{bm}
\usepackage{braket}
\usepackage{xcolor} 
\usepackage[normalsize]{subfigure}     
\usepackage{amsthm}
\usepackage{hyperref}
\usepackage{CJK} 

\newtheorem{theorem}{Theorem}
\newtheorem{definition}[theorem]{Definition}

\newcommand{\CFT}{{\textsl{\tiny CFT}}}
\newcommand{\calH}{\mathcal{H}}

\newcommand{\bbV}{\mathbb{V}}

\newcommand{\bfP}{\mathbf{P}}

\DeclareMathOperator{\Tr}{Tr}
\DeclareMathOperator{\sgn}{sgn}

\definecolor{dkgreen}{rgb}{0,0.6,0}

\definecolor{purple}{rgb}{0.5,0,0.5}

\begin{document}

\title{A universal tripartite entanglement signature of ungappable edge states}
\date{\today}
\author{Karthik Siva$^*$}
\affiliation{Department of Physics, University of California, Berkeley, CA 94720, USA}
\altaffiliation{These authors contributed equally to this work}
\author{Yijian Zou$^*$}
\affiliation{Stanford Institute for Theoretical Physics, Stanford University, Palo Alto, CA 94305, USA}
\altaffiliation{These authors contributed equally to this work}
\author{Tomohiro Soejima}%
\affiliation{Department of Physics, University of California, Berkeley, CA 94720, USA}
\author{Roger S. K. Mong}
\affiliation{Department of Physics and Astronomy, University of Pittsburgh, Pittsburgh, PA 15260, USA}
\author{Michael P. Zaletel}
\affiliation{Department of Physics, University of California, Berkeley, CA 94720, USA}
	\affiliation{Materials Sciences Division, Lawrence Berkeley National Laboratory, Berkeley, California 94720, USA
}

\begin{abstract}
Gapped two-dimensional topological phases can feature ungappable edge states which are robust even in the absence of  protecting symmetries. 
In this work we show that a multipartite entanglement measure recently proposed in the context of holography, the Markov gap, provides a universal diagnostic of  ungappable edge states.
Defined  as a difference of the reflected entropy and mutual information $h(A:B) = S_R(A:B) - I(A:B)$ between two parties, we argue that for $A,B$ being adjacent subregions in the bulk, $h=\frac{c_+}{3}\log 2$, where $c_+$ is the minimal total central charge of the boundary theory. As evidence, we prove that $h=0$ for string-net models, and numerically verify that $h=\frac{|C|}{3}\log 2$ for a Chern-$C$ insulator. 
Our work  establishes a unique bulk entanglement criteria for the presence of a conformal field theory on the boundary.

\end{abstract}

\maketitle

Long-range entangled topological phases are characterized by a pattern of ground state entanglement which cannot be adiabatically transformed into a product state~\cite{Wen2017}.
Among  two-dimensional (2D) topological orders (TOs) we may distinguish between two types.
``Ungappable'' TOs, such as the integer and fractional quantum Hall effects, have irremovable gapless edge states~\cite{Thouless:Pump83,Wen1991,HatsugaiIQHE93}, even in the absence of protecting symmetries.
One mechanism for such behavior is a mismatch between the number of left and right movers, $c_- = c_R - c_L$, of the conformal field theory (CFT)  governing the edge; although certain non-chiral theories with fractionalized excitations  are also ungappable~\cite{Kitaev2006,levin2013, lan2015gapped}.
On the other hand, ``gappable'' TOs such as the toric code \cite{Kitaev2003} and string-net states~\cite{Levin2005} admit a gapped boundary theory for a suitable choice of edge Hamiltonian~\cite{kitaev2012}.
It has been shown that a TO is gappable if \emph{and only if} it admits a string-net representation \cite{kitaev2012,lin2014generalizations, freed2021gapped}.

While topological order has traditionally been probed via its excitations (e.g.\ edge states and fractionalized quasiparticles), more recently quantum information measures have been used as a method for detecting the pattern of long-range entanglement in the ground state itself.
Previous work has largely focused on bipartite entanglement. In particular the topological entanglement entropy (TEE) \cite{Kitaev2006,Levin2006,Zhang2012}, which is a linear combination of the entanglement entropy of different subregions, measures the total quantum dimension of the anyonic excitations.
However, the TEE does not distinguish between ungappable and gappable TO.
Indeed the integer quantum Hall effect is an ungappable TO with vanishing TEE.
In order to distinguish \emph{chiral} ungappable TO from gappable TO, one may look at the entanglement spectrum, which has the same anomalies as the physical edge \cite{Kitaev2006, Li2008}. 
On a translationally invariant cylinder it was argued that the entanglement spectrum encodes $c_- \bmod 24$ \cite{Zaletel2013,Tu2013}.
However, this approach relies on translation symmetry and does not detect non-chiral ungappable TOs.
A quantitative bulk entanglement criteria for ungappable TO is thus still lacking.

In this work we provide such a measure by going beyond bipartite entanglement  and considering a multipartite entanglement measure $h(A:B)$ recently referred to as the ``Markov gap'' \cite{hayden2021}.  This quantity was first discussed in the context of 1+1D CFTs and holography~\cite{Dutta2019,Akers2020,Zou2021,hayden2021}. After defining a procedure which eliminates the non-universal short-distance contribution, we argue that the remainder takes the universal value
\begin{align} h_{\mathrm{IR}}=\frac{c_+}{3}\log 2 , \label{eq:hIR_c} \end{align}
where $c_+ = c_L + c_R$ is the minimal total central charge of a \emph{single} edge~%
    \footnote{For a 1D non-anomalous critical chain, the total is $c_+ = 2c$, where $c$ is the central charge (as conventionally used) of the 1D system, because $c = c_L = c_R$.} of the TOs boundary CFT.
To give evidence for our conjecture we first prove that $h=0$ for string-net states, consistent with their gappable edge.
Second, we numerically compute $h$ for integer quantum Hall states  and find excellent agreement with Eq.~\eqref{eq:hIR_c}.
Compared with previous work, our method has two merits. First, it relies only on the reduced density matrix of a subregion in the bulk and does not require translation or other symmetries.
Second, it is a quantitative measure which determines the central charge of the boundary CFT, with finite-size corrections which appear to converge exponentially. 
Our work thus establishes a quantitative bulk entanglement criteria distinguishing between gappable and ungappable TO.

\textit{Markov gap $h(A:B)$} --- We start by defining the multipartite entanglement measure $h$ for a quantum state. Given a pure state $|\psi\rangle_{ABC}$ tripartitioned into $A,B$ and $C$, the reduced density matrix on $AB$ is given by $\rho_{AB} = \Tr_C |\psi\rangle\langle\psi|$. One purification of $\rho_{AB}$, known as the canonical purification, is given by the square root of the density matrix taken as a state $|\sqrt{\rho}\rangle_{ABA^{*}B^{*}}$ in $\calH_{A}\otimes \calH_{B} \otimes \calH^{*}_{A} \otimes \calH^{*}_{B}$. The reflected entropy $S_R(A:B)$ is given by the entanglement entropy $AA^{*}$ in the canonical purification, $S_R(A:B) = S_{AA^{*}}(|\sqrt{\rho}\rangle_{ABA^{*}B^{*}})$~\cite{Dutta2019}.
The Markov gap $h(A:B)$ is then defined as
\begin{equation}
    h(A:B) = S_R(A:B)-I(A:B) \geq 0,
\end{equation}
where $I(A:B)$ is the mutual information between $A$ and $B$. As shown in Ref.~\onlinecite{Zou2021}, $h$ is a nonnegative quantity that vanishes if and only if the state $|\psi\rangle_{ABC}$ has an algebraic form given by a sum of triangle states (SOTS).
\begin{definition}
A pure state $\ket{\psi}_{ABC}$ is a SOTS if for each local Hilbert space $\calH_{\alpha}$ ($\alpha \in \{ A, B, C \} $) there exists a decomposition $\calH_\alpha = \bigoplus_j \calH_{\alpha^{j}_L}\otimes \calH_{\alpha^{j}_R}$ such that
\begin{equation}
\label{eq:SOTS}
    \ket{\psi}_{ABC}=\sum_{j} \sqrt{p_j} \ket{\psi_{j}}_{A^{j}_RB^{j}_L}\ket{\psi_{j}}_{B^{j}_RC^{j}_L}\ket{\psi_{j}}_{C^{j}_RA^{j}_L},
\end{equation}
and $\sum_{j}p_j = 1$.
\end{definition}
Roughly speaking, a SOTS only contains bipartite entanglement and Greenberger–Horne–Zeilinger (GHZ) type of entanglement. 
Therefore, nonvanishing $h(A:B)$ indicates entanglement across the three subregions beyond the GHZ type.

\begin{figure}
    \centering
    \subfigure[]{\includegraphics[width=0.4\linewidth]{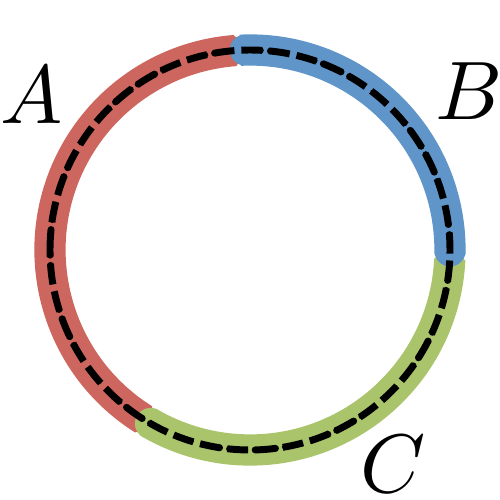}}
    \qquad
    \subfigure[]{\includegraphics[width=0.4\linewidth]{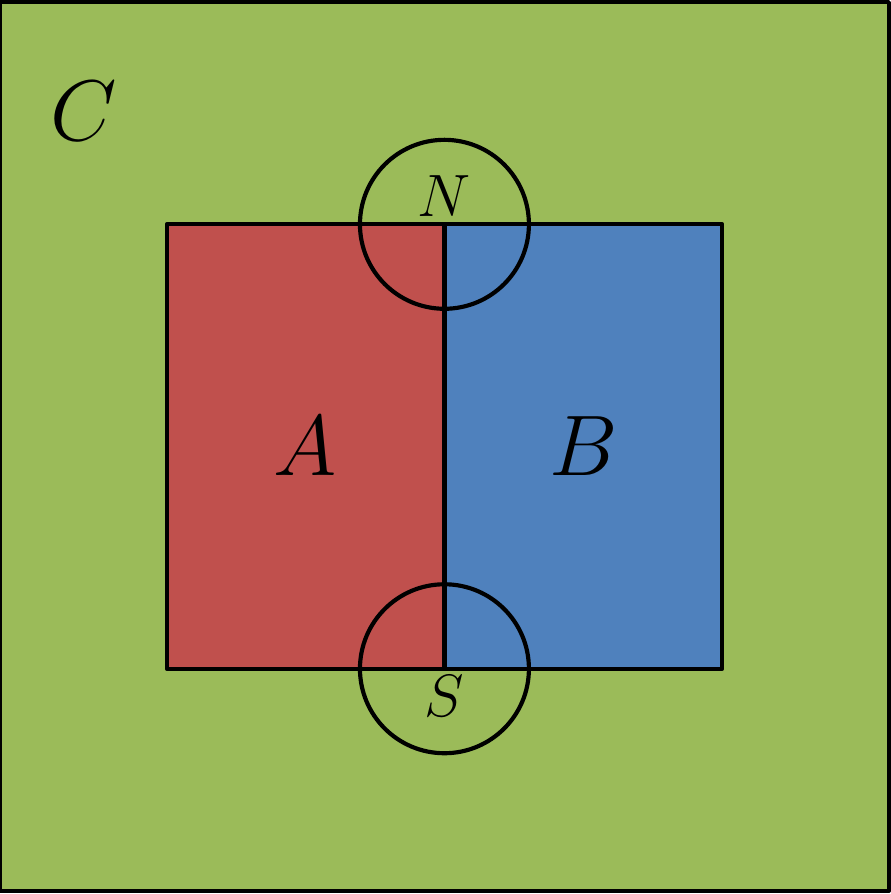}}
    \caption{(a) Tripartition of a 1D system on the circle. (b) A 2D system on a square lattice.}
    \label{fig:tripartition}
\end{figure}

\textit{Tripartition for 1D and 2D systems} --- In Fig.~\ref{fig:tripartition} we show the tripartition that is considered in this paper. For a one-dimensional system, we choose $A,B,C$ to be adjacent intervals.
It has been shown \cite{Zou2021,Dutta2019} that the ground state of a gapped system has $h=0$ and the ground state of a gapless system has $h=h^{\CFT} \equiv \frac{c}{3}\log 2$ , where $c$ is the central charge of the CFT. 

\begin{figure}
    \centering
    \includegraphics[width = 0.9\linewidth]{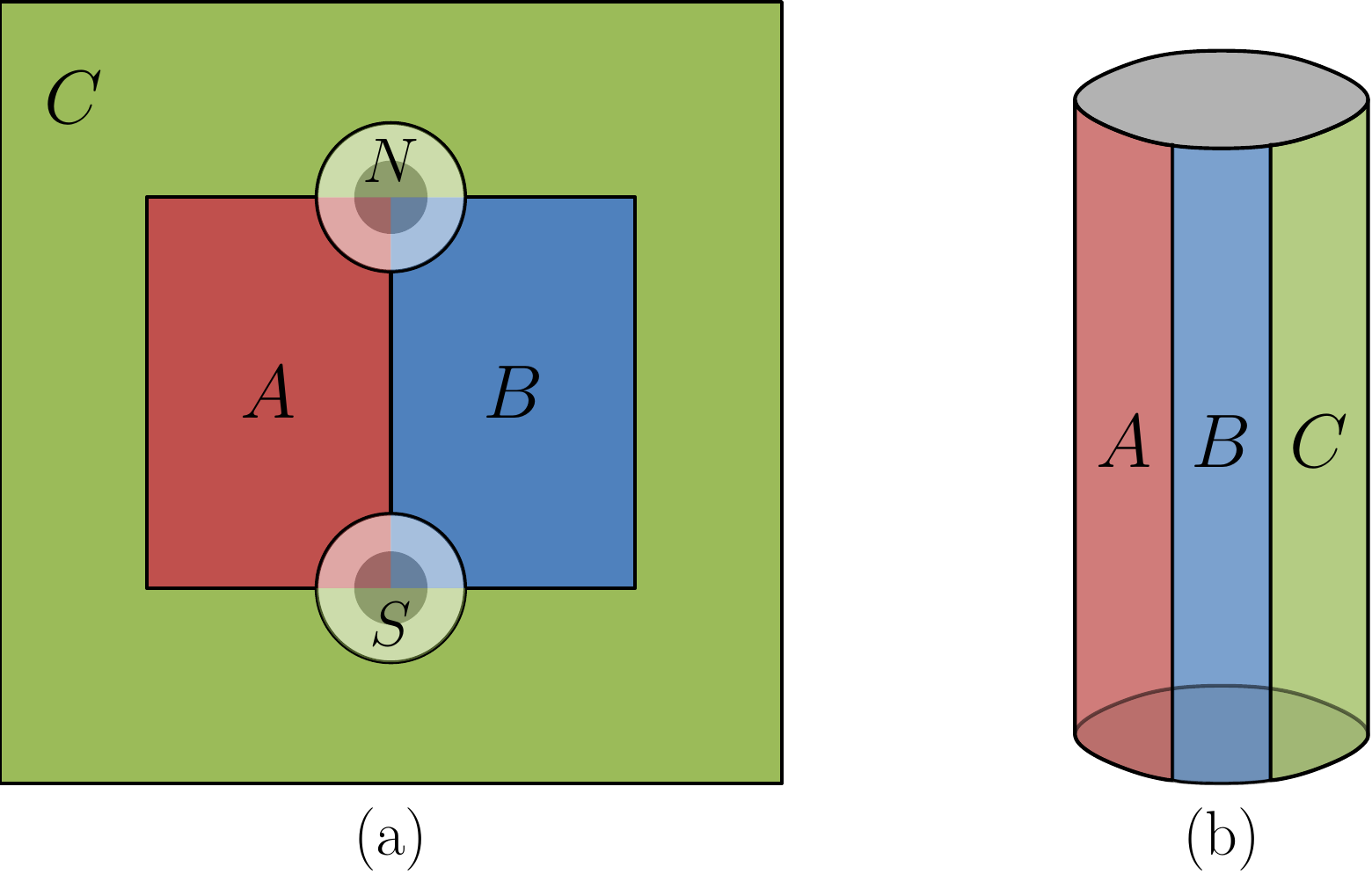}
    \caption{(a) The tripartition of a two dimensional system with punctures illustrated by shaded regions. There are left- and right-moving modes on the boundary of the shaded regions. Starting from the geometry in Fig.~\ref{fig:tripartition}(b), we act with unitary disentanglers on the larger circular regions $N$ and $S$ (including the shaded interiors). Intuitively, in systems with a gapped bulk, such a unitary can be chosen to turn the degrees of freedom in the shaded interior into a product state. (b) An open cylinder with a tripartition where the three parties are strips that connect partitions of the two circles. Viewing the product state regions as punctures, the two geometries are topologically equivalent and can be deformed into each other using finite-depth local unitaries.}
    \label{fig:smoother}
\end{figure}

Now we consider a two-dimensional lattice with the tripartition given in Fig.~\ref{fig:tripartition}. In contrast to the one dimensional case, there are two trisection points $N$ and $S$ where the three regions meet. The trisections can contribute a lattice-scale \emph{non-universal} contribution to $h(A:B)$.
Intuitively, UV physics can dress the trisection with an entangled tripartite state, which can contribute a finite $h$. 
There are two ways around this non-universal contribution. First, one may consider a modified geometry in which a disk is removed from each trisection, so that the system becomes topologically equivalent to an open cylinder (Fig.~\ref{fig:smoother}).
As will become clear, this approach does work, but it creates additional edges in the bulk.
However, since we aim to demonstrate that $h$ can be made a universal quantity  \emph{purely from the bulk ground state}, we instead develop a method for the disk geometry.

Instead of the ground state $|\psi\rangle_{ABC}$, we consider the space of ``smoothed'' states $U_{N} U_{S}|\psi\rangle_{ABC}$, where $U_{N/S}$ is a unitary supported on a circle of radius $R$ centered at $N/S$. We define the bulk entanglement quantity $h_R$ at length scale $R$ as
\begin{equation}
\label{eq:def_h_2D}
    h_R \equiv \min_{U_N,U_S} h(A:B).
\end{equation}
We then define $h_{\mathrm{IR}} = \lim_{R \to \infty} h_R$, where the limit is such that $A, B, C$ must all be kept large in comparison with $R$. In practice, we will argue $h_R$ converges exponentially quickly at a rate which is determined by some length scale $\xi$ of the bulk ground state.
The main result of this paper is that $h_{\mathrm{IR}}=\frac{c_+}{3} \log 2$ for a 2D system, where $c_+$ is the minimal central charge of the boundary theory. For a gapped theory $c_+=0$, and for a ungappable theory $c_+\geq \frac{1}{2}$.
Note that by construction $h \geq h_{\mathrm{IR}} = \frac{c_{+}}{3}\log 2$, so our result may also be interpreted as a lower-bound on the bare value of $h$.


\textit{Argument for universal $h_\mathrm{IR}$} --- Here we make an intuitive argument for the main result. Suppose $U_{N/S}$ (of radius $R$) are chosen to transform a subregion of radius $R' < R$ centered on each trisection into a product state.  This can always be done if we allow for a buffer of width $\xi < R - R'$, where $\xi$ is related to a correlation length. 
Physically, for example, this operation can be accomplished by adiabatically turning on a topologically trivial mass term around the trisection.
We then view the product state subregions as ``punctures'' [shaded in grey in Fig.~\ref{fig:smoother}(a)].
If $C$ is one-point compactified at infinity, the geometry is topologically equivalent to Fig.~\ref{fig:smoother}(b), where $A,B,C$ are strips winding around an open cylinder. For an ungappable TO, there are edge modes on the top and bottom circles of the cylinder. As the bulk is gapped, we may compress the vertical direction such that the top and bottom meet, reducing to the one-dimensional geometry of Fig.~\ref{fig:tripartition}(a). The theory on the circle is the full boundary CFT of the  TO by combining left and right moving modes on the two boundaries of the cylinder. As $h(A:B)$ is invariant under local unitary operations, we expect that $h(A:B) = h^{\CFT} = \frac{c_+}{3}\log 2$ with the disentangler applied. As $h_R$ is the minimum over $U_{N},U_S$, we may take $h^\CFT$ given by the particular choice of disentangler in the thought experiment above to provide an upper bound on $h_R$. We expect that the disentangler gives the optimal $h$ as the contribution from the edge modes are ungappable by local perturbations. Thus $h_R = h^{\CFT}$ if $R\gg\xi$.

If the edges are instead gapped, after compressing the cylinder along the vertical direction, the remaining state can be described by a matrix product state with finite bond dimension. For such states, it was proved in Ref.~\onlinecite{Zou2021} that $h(A:B)=0$. We may think of the smoothers on the disk as ensuring that when the trisection is mapped to the edge of a cylinder, a local non-universal tripartite entangled state does not become global.

Returning to the disk, since the reduced density matrix of $A,B$ plus the range of smoothers only measures local correlations, it is not affected by the topology of the whole system. The quantity $h_R$ defined in Eq.~\eqref{eq:def_h_2D} is thus universal given that the length scales of $A,B,C$ much larger than $R$ and $R \gg \xi$. Below we confirm the statement by showing that (i) $h=0$ for string-net states, which are commonly believed to give a complete classification of topological orders with gappable edges in 2D, and that (ii) $h = \frac{c}{3}\log 2$ for a stack of Chern insulators, where $c$ is the minimal central charge of the boundary CFT. 

\textit{Vanishing $h$ for gappable topological order} --- We consider a tripartition as in Fig.~\ref{fig:tripartition} of the string-net liquids introduced in Ref.~\onlinecite{Levin2005} which (can be generalized to) characterize the fixed-point (zero correlation length) wavefunctions of all topological orders with gappable edges in two dimensions.
For string-net states $\xi = 0$, so we do not need the smoothers at all, and we are able to show explicitly that they have a SOTS structure. We review the formalism of string-net liquids and present the general proof in the Supplementary Material but illustrate the argument here for the simple case of the toric code~\cite{Kitaev2003}.

\begin{figure}
    \centering
    \includegraphics[width=1.0\linewidth]{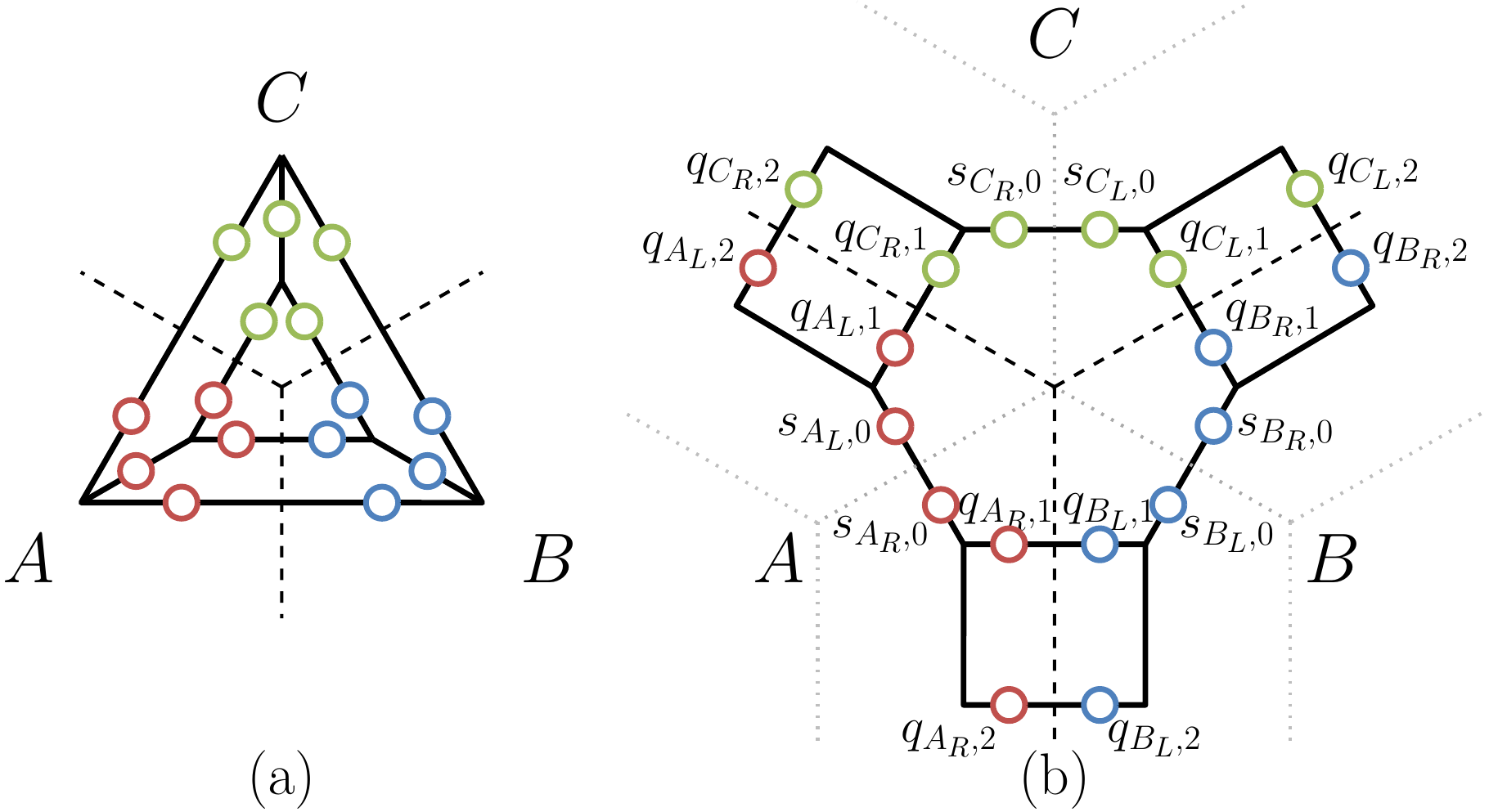}
    \caption{(a) Toric code on a mesh covering the surface of a sphere. Degrees of freedom (0 or 1, denoted by circles) live on links, and the ground state is a uniform superposition over configurations of 0's and 1's satisfying the constraint that at each vertex, an even number of 1's must meet. (b) The configuration in (a) may be reduced to this one using local unitaries acting on $A,B$ and $C$. }
    \label{fig:toric_sphere}
\end{figure}

Consider the toric code on a trivalent lattice in Fig.~\ref{fig:toric_sphere}.  The degrees of freedom are located on the links of the lattice, and a two-dimensional Hilbert space is associated to each link. Let $\{\ket{0}, \ket{1}\}$ label a basis for each space. The wavefunction is the ground state of the following Hamiltonian:
\begin{align}
    H = -\sum_v A_v &- \sum_p B_p,\\
    A_v = \prod_{i\in v} \sigma^z_i,&\quad B_p = \prod_{i} \sigma^x_i,
\end{align}
where $v$ ($p$) labels vertices (plaquettes) of the lattice, and $A_v$ ($B_p$) is supported on the links connected to vertex (plaquette) $v$ ($p$). All $A_v$'s commute with all $B_p$'s, and the ground state is the mutual $+1$ eigenstate of all of the operators. We can interpret the term $A_v$ as enforcing the constraint that at vertex $v$, the sum of three links must be $0 \bmod 2$. The term $B_p$ then interchanges different configurations of $0$'s and $1$'s, and the global ground state is a uniform superposition over all configurations satisfying the constraints on the vertices.

The lattice shown in Fig.~\ref{fig:toric_sphere} (a) may be viewed as covering the surface of a sphere tripartitioned according to the dashed black lines. Regions $A$ and $B$ are adjacent, and $C$ is complementary, and degrees of freedom on links straddling the partitions are doubled. Since this wavefunction has zero correlation length, we may employ such a minimal representation of the toric code. As in Ref.~\onlinecite{Levin2006} we may reduce each region, but because each has a boundary it can only be reduced to a tree-like diagram as in Fig.~\ref{fig:toric_sphere} (b). 

We now analyze the tripartite wavefunction in Fig.~\ref{fig:toric_sphere} (b). The doubling on links implies that $q_{A_L,i}=q_{C_R,i}$, $q_{C_L,i}=q_{B_R,i}$, etc. as well as $s_{A_L,0}= s_{A_R,0}$,$s_{B_L,0}= s_{B_R,0}$ etc. Next, the graphical rules describing the relations between string configurations in Ref.~\cite{Levin2005} require that $s_{A_L,0} = s_{C_R,0} =  s_{C_L,0} = s_{B_R,0} = s_{B_L,0} = s_{A_R,0}$. Let $s$ denote the value of these central degrees of freedom. The total wavefunction may be organized as a sum over the central $s$, and the value of each $q_{\alpha,2}$ is set by the fusion of $s$ and $q_{\alpha,1}$ for $\alpha \in \{A_L, A_R, B_L,\dots\}$.
Define
\begin{align}
\label{eq:bell}
    \ket{AB(s)} &\equiv \frac{1}{\sqrt{2}}\ket{s}_{A_R,0}\ket{s}_{B_L,0} \\ \nonumber &\quad\otimes\sum_{q=\{0,1\}}\ket{q}_{A_R,1}\ket{q}_{B_L,1}\ket{q\oplus s}_{A_R,2}\ket{q\oplus s}_{B_L,2}
\end{align}
and similarly define $\ket{BC(s)}$ and $\ket{CA(s)}$. For each value of $s$, this state in Eq.~\eqref{eq:bell} is essentially a Bell pair between $A_R$ and $B_L$, and $\braket{AB(1)|AB(0)}=0$. The ground state $\ket{\psi_\text{gs}}$ may then be written as 
\begin{equation}
    \ket{\psi_\text{gs}} = \sum_{s=\{0,1\}} \frac{1}{\sqrt{2}} \ket{AB(s)}\otimes \ket{BC(s)} \otimes \ket{CA(s)}
\end{equation}

This wavefunction manifestly satisfies the SOTS form in Eq.~\eqref{eq:SOTS}, with the factorization into $L$ and $R$ Hilbert spaces on each region as indicated by the dotted gray line in Fig.~\ref{fig:toric_sphere} (b), and we can therefore conclude that $h(A:B)=0$ for toric code. In the Supplemental Material, we apply this approach more generally to other string-net wavefunctions and find that they, too, may be written as a SOTS as in Eq.~\ref{eq:SOTS}.

\textit{Universal $h$ for stacked Chern insulators} --- The simplest chiral topological order is the Chern insulator, where the minimal central charge $c_+$ on the boundary is given by the magnitude of the Chern number $|C|$ of the bulk \cite{TKNN:82,ASS:Homotopy:83,Haldane:QHE:88}. The Chern insulator can be realized on a lattice by a tight-binding model coupled to an external magnetic field $B$ \cite{Hofstadter1976}. We consider the Hamiltonian given by
\begin{equation}
    H(B)= -t\sum_{\vec{x},\vec{a}} (c^{\dagger}_{\vec{x}} e^{-i \vec{a} \cdot \vec{A}(\vec{x})} c_{\vec{x}+\vec{a}} + h.c.)+\mu \sum_{\vec{x}} c^{\dagger}_{\vec{x}} c_{\vec{x}}
\end{equation}
where $\vec{a}$ runs over lattice vectors and $\vec{A}$ is the vector potential which equals $\vec{A} = (0,B x_1)$ for the square lattice in the Landau gauge. The Chern number of the system is a sum of the Chern numbers of individual bands that are filled. We consider the lowest band for $B=\pm \pi/2$, which has $C=\pm 1$ and the lowest two bands for $B=\pi/3$, which both have $C=1$. A topological insulator \cite{Kane2005,bernevig2006} can be constructed by stacking two layers of the Chern insulator with $C=1$ and $C=-1$. The topological insulator is an example of symmetry-protected topological (SPT) phase where the edge modes are protected by the time reversal (TR) symmetry~\cite{KaneMele:Z2:2005}. The minimal central charge on (both) the boundaries is $c_+=0$ if TR is broken and $c_+=2$ if TR is preserved.

The model is quadratic in the fermionic variables and the entanglement quantities can be computed by the standard covariance matrix techniques \cite{Peschel2009,Bueno2020}. In order to obtain $h_R$ in Eq.~\eqref{eq:def_h_2D}, we restrict the generators of the smoothers $U_{N/S}$ to be quadratic in the fermionic variables. When the edge modes are protected by TR symmetry, we further demand the $U_{N/S}$ are generated by a TR-invariant flow.
The smoothers are optimized with a gradient optimization, where the gradient can be computed from the covariance matrix. We compute the optimized $h$ for different disentangler sizes up to $R = 6$ and different subsystem sizes up to $L_A = L_B = 24$. We find that $h$ is independent of the sizes once $\xi\ll R \ll L_A,L_B$ and that $h=\frac{c_+}{3}\log 2$, where $c_+$ is the minimal central charge of the edge modes. The numerical result is shown in Fig.~\ref{fig:h_hof} and Tab.~\ref{tab:h}. Further details can be found in the Supplemental Material.

\begin{figure}
    \centering
    \includegraphics[width=0.9\linewidth]{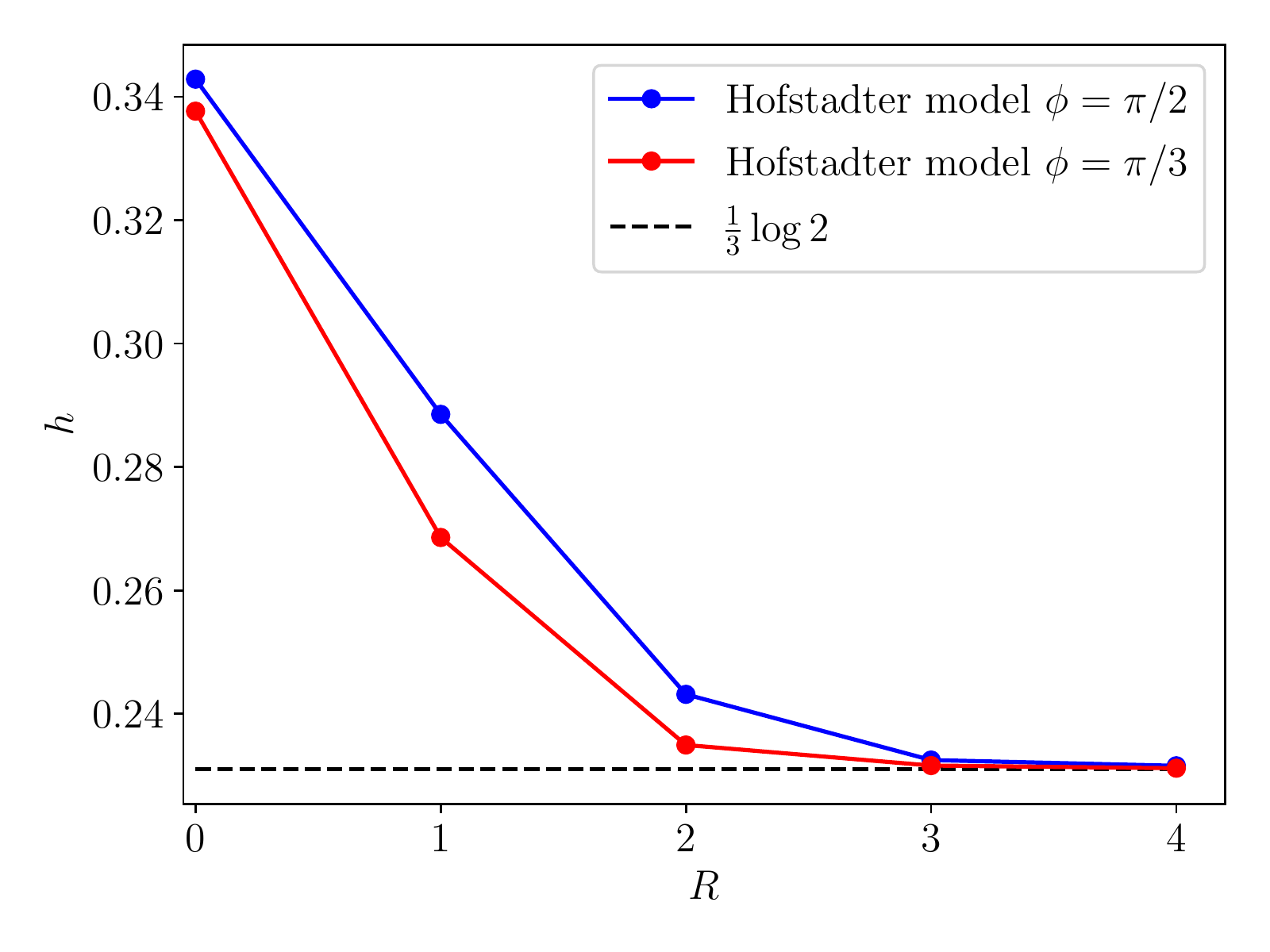}
    \caption{$h$ for the Hofstadter model with Chern number $C=1$.  The dashed line denote the theoretical value $\frac{1}{3}\log2$.}
    \label{fig:h_hof}
\end{figure}

\begin{table}[h]
    \centering
    \begin{tabular}{|c|c|c|c|}
    \hline
       Bulk TO  & $c_+$ & $h^{\CFT}$ & $h_R$    \\ \hline
         String net & 0 & 0 & 0 \\ \hline
         $B=\pi/2$, lowest band & 1 & 0.2310 & 0.2316 \\ \hline
         $B=\pi/3$, lowest band & 1 & 0.2310 & 0.2312 \\ \hline
         $B=\pi/3$, lowest two bands & 2 & 0.4621 & 0.4641 \\ \hline
         Topological insulator (TR preserved) & 2 & 0.4621 & 0.4632 \\  \hline
         Topological insulator (TR broken) & 0 & 0 & 0.0014 \\ \hline
    \end{tabular}
    \caption{$h$ as defined in Eq.~\eqref{eq:def_h_2D} for different topological order. The second column shows the central charge for the edge modes. The third column shows the theoretical $h^{\CFT} \equiv \frac{c_+}{3}\log 2$. The fourth column shows the values of $h_R$ for different lattice models. $h_R$ iscomputed analytically for string nets and optimzed numerically for other models. The smoother range is $R=6$ for $B=\pi/3$ with lowest two bands filled and $R=4$ for other models.}
    \label{tab:h}
\end{table}


\textit{Discussion} --- In this paper we have established a bulk multipartite entanglement quantity $h_{\mathrm{IR}}$ for two-dimensional topologically ordered systems. We have shown that $h_{\mathrm{IR}}=\frac{c_+}{3}\log 2$, where $c_+$ is the minimal central charge of the boundary CFT. One  numerically irksome feature of the definition is the use of disentanglers to remove short-distance entanglement at the trisection points. It would be interesting if instead a ``subtraction scheme'', as in the TEE, could be devised.

One may naturally wonder whether the result is sensitive to the form of the disentangler. We first note that if the disentanglers were restricted entirely in $AB$, then we would no longer obtain $h_{\mathrm{IR}}=h^{\CFT}$, as we are no longer able to puncture a hole near the trisections.
Next, as shown in the Supplemental Material, we find $h=0$ for all models if the unitaries are allowed to entangle the degrees of freedom of the two tripartitions (i.e., when acting with a joint $U_{NS}$).
This is expected as it allows the left-moving modes to hybridize with the right-moving modes such that they can be removed simultaneously.
We also found that when the bulk is in an SPT phase, we obtain different $h_{\mathrm{IR}}$ depending on whether the disentanglers respect or break the symmetry. Therefore, in addition to distinguishing between gappable and ungappable long-range entanglement,  $h$ may also be used to detect short-range entangled symmetry protected topological order. 


\emph{Note added} --- Near the completion of this work, we became aware of independent work by Liu et~al.~\cite{Liu2021}, which also considers the reflected entropy of 2D states, and Kim et~al.~\cite{kim2021}, which proposes a distinct entanglement measure for detecting the chiral central charge of a 2D topological state.

We are grateful for illuminating conversations with David Aasen and Shinsei Ryu and are indebted to Brian Swingle for the suggestion to consider the topological insulator.
YZ is supported by the Q-FARM fellowship at Stanford University. Part of the research was done at Perimeter Institute; research at Perimeter Institute
is supported by the Government of Canada through the
Department of Innovation, Science and Economic Development Canada and by the Province of Ontario through the Ministry of Research, Innovation and Science.
KS was supported by the  U.S. Department of Energy, Office of Science, National Quantum Information Science Research Centers, Quantum Systems Accelerator.
TS was supported by the Masason foundation.
RM was supported by the National Science Foundation grant No.~DMR-1848336.

\bibliography{main}

\clearpage
\appendix
\onecolumngrid
\section{The covariance matrix formalism}
In the section we first review the covariance matrix and then derive the algorithms to compute the reflected entropy. Finally we derive the gradient method to optimizing $h(A:B)=S_R(A:B)-I(A:B)$ using disentanglers.
\subsection{Covariance matrix}
We consider a lattice consisting of $N$ complex fermion modes whose creation operators $c^{\dagger}_i$ and annihilation operators $c_i$ satisfy the anticommutation relation
\begin{equation}
    \{c_i, c^{\dagger}_j\}=\delta_{ij},~~~ \{c_i, c_j\}=\{c^{\dagger}_i, c^{\dagger}_j\}=0.
\end{equation}
A Gaussian state is a state whose density matrix can be represented as
\begin{equation}
    \rho = \frac{1}{Z} \exp\left(-\sum_{i,j=1}^{N}c^{\dagger}_i h_{ij} c_j\right),
\end{equation}
where $h$ is a $N\times N$ Hermitian matrix and 
\begin{equation}
    Z=\Tr \exp\left(-\sum_{i,j=1}^{N}c^{\dagger}_i h_{ij} c_j\right)
\end{equation}
is the partition function. Let $U$ be a unitary matrix that diagonalizes $h$,
\begin{equation}
\label{eq:entang_modes}
    UhU^{\dagger} = \mathrm{diag}(\epsilon_k),
\end{equation}
where $\epsilon_k$ denotes the eigenvalues, then
\begin{equation}
\label{eq:rho_eigenbasis}
    \rho = \frac{1}{Z} \prod_{k=1}^N \exp\left(-\epsilon_k\tilde{c}^{\dagger}_k \tilde{c}_k \right),
\end{equation}
where
\begin{equation}
\label{eq:eigenbasis}
    \tilde{c}_k = \sum_{l=1}^N U_{kl} c_l.
\end{equation}
The partition function can be represented as 
\begin{equation}
    Z \equiv \prod_{k=1}^{N} Z_k = \prod_{k=1}^N \left(1+e^{-\epsilon_k}\right).
\end{equation}
A Gaussian state can be characterized by the covariance matrix 
\begin{equation}
    C_{ij} \equiv \langle c^{\dagger}_i c_j \rangle = \Tr(\rho c^{\dagger}_i c_j).
\end{equation}
By definition, the covariance matrix is Hermitian, $C_{ij}=C^{*}_{ji}$. The covariance matrix can be diagonalized by the same unitary $U$,
\begin{equation}
    UCU^{\dagger} = \mathrm{diag}(\langle \tilde{c}^{\dagger}_k \tilde{c}_k \rangle),
\end{equation}
where
\begin{equation}
\label{eq:FD_distribution}
    \langle \tilde{c}^{\dagger}_k \tilde{c}_k \rangle=\frac{1}{1+e^{\epsilon_k}}
\end{equation}
is the Fermi-Dirac distribution. Combining Eqs.~\eqref{eq:entang_modes} and \eqref{eq:FD_distribution} we obtain the matrix equation
\begin{equation}
    h = \log\left(\frac{I-C}{C}\right).
\end{equation}

\subsection{Entanglement quantities from the covariance matrix}
One may compute entanglement quantities of a Gaussian state using the covariance matrix. Consider a pure state that occupies $n$ orbits with wavefunctions $\psi_{ij}$, where $i=1,2,\cdots N$ labels the sites and $j=1,2,\cdots n$ labels the orbits,
\begin{equation}
    |\psi\rangle = \prod_{j=1}^{n} \left(\sum_{j=1}^N\psi_{ij} c^{\dagger}_i\right)|0\rangle.
\end{equation}
The covariance matrix of the state is given by
\begin{equation}
    C=\psi\psi^{\dagger}.
\end{equation}
Note that for a pure state $C^2=C$. Given a subsystem $A$ of the lattice, the covariance matrix of the subsystem is given by $C_A$, the submatrix of $C$ restricted to the rows and columns corresponding the subsystem $A$. The reduced density matrix of the subsystem $A$ is then
\begin{equation}
\label{eq:rho_R}
        \rho_A = \frac{1}{Z_A} \exp\left(-c^{\dagger} h_A c\right),
\end{equation}
where
\begin{equation}
\label{eq:h_R}
    h_A= \log\left(\frac{I-C_A}{C_A}\right).
\end{equation}
The entanglement entropy 
\begin{equation}
    S(A)=-\Tr(\rho_A \log \rho_A).
\end{equation}
Using Eqs.~\eqref{eq:rho_R} and \eqref{eq:h_R} it is straightforward to show that
\begin{equation}
\label{eq:fermionEE}
    S(A) = -\Tr(C_A\log C_A + (I-C_A)\log(I-C_A)).
\end{equation}
The mutual information between two subsystems $A,B$ is given by 
\begin{equation}
    I(A:B)=S(A)+S(B)-S(AB),
\end{equation}
where each term can be obtained by Eq.~\eqref{eq:fermionEE}. 

Now we compute the reflected entropy $S_R(A:B)$ using the covariance matrix. Given a density matrix $\rho_{AB}$, the canonical purification is  $\mathcal{P}|\sqrt{\rho_{AB}}\rangle$, where $\mathcal{P}$ is an isomorphism that transforms a bra state in $ \calH^{*}_{AB}$ into a ket state in $\calH_{A'B'}$ and $\calH_{A'B'}\cong \calH_{AB}$ is the auxiliary space for the purification. It is clear that $\mathcal{P}$ is an antiunitary operator. For a free fermion state, we may choose the antiunitary operator to coincide with the particle-hole transformation on the auxiliary system,
\begin{equation}
\label{eq:PHtransform}
    \mathcal{P}(\langle \{n_i\}|)= |\{1-n_i\}\rangle,
\end{equation}
where $|\{n_i\}\rangle$ is the simultaneous eigenstate of the number operator $c^{\dagger}_ic_i$ with eigenvalue $n_i$ on each site.
The reflected entropy is
\begin{equation}
    S_R(A:B)=S_{AA'}(\mathcal{P}|\sqrt{\rho_{AB}}\rangle).
\end{equation}
The canonical purification of a Gaussian state is still a Gaussian state. Therefore, in order to compute the reflected entropy one only needs the covariance matrix $C^{R}$ of the canonical purification $\mathcal{P}|\sqrt{\rho_{AB}}\rangle$. Using Eq.~\eqref{eq:rho_eigenbasis} and the particle hole transormation Eq.~\eqref{eq:PHtransform} we obtain
\begin{equation}
    \mathcal{P}|\sqrt{\rho_{AB}}\rangle = \prod_{k=1}^N \left( \frac{1}{\sqrt{Z_k}}\left(|0_{k}\rangle|1_{\tilde{k}}\rangle+ e^{-\epsilon_k/2}|1_k\rangle|0_{\tilde{k}}\rangle\right) \right).
\end{equation}
The correlation functions are
\begin{eqnarray}
        \langle \tilde{c}^{\dagger}_k \tilde{c}_k \rangle &=& \frac{1}{1+e^{\epsilon_k}},      \\    
        \langle \tilde{c}^{\dagger}_{k'} \tilde{c}_{k'} \rangle &=& \frac{e^{\epsilon_k}}{1+e^{\epsilon_k}}=1- \langle \tilde{c}^{\dagger}_k \tilde{c}_k \rangle\\
        \langle \tilde{c}^{\dagger}_{k} \tilde{c}_{k'} \rangle &=& \frac{e^{\epsilon_k/2}}{1+e^{\epsilon_k}}= \sqrt{\langle \tilde{c}^{\dagger}_k \tilde{c}_k \rangle\langle \tilde{c}^{\dagger}_{k'} \tilde{c}_{k'} \rangle}
\end{eqnarray}
Undoing the unitary transformation Eq.~\eqref{eq:eigenbasis}, one finds the correlation matrix
\begin{equation}
    C^{R}=
    \begin{bmatrix}
    C_{AB} & \sqrt{C_{AB}(I-C_{AB})} \\
    \sqrt{C_{AB}(I-C_{AB})} & I-C_{AB}
    \end{bmatrix},
\end{equation}
where the off-diagonal block represents correlations between the region $AB$ and its reflection $A'B'$. Note also that $(C^{R})^2=C^{R}$, consistent with the fact that the canonical purification is a pure state. Finally the reflected entropy can be computed analogous to Eq.~\eqref{eq:fermionEE},
\begin{equation}
\label{eq:fermionSR}
    S_R(A:B)=-\Tr(C^{R}_{AA'}\log C^{R}_{AA'} + (I-C^{R}_{AA'})\log(I-C^{R}_{AA'})),
\end{equation}
where $C^{R}_{AA'}$ is the submatrix of $C_R$ restricted to rows and columns corrsponding to $A$ and $A'$.
\subsection{Computing the disentangler}
Given a pure state $|\psi\rangle_{ABC}\rangle$ and a subregion $S\in{ABC}$ (not necessarily in $AB$), we aim to find the minimization algorithm for
\begin{equation}
    h_S = \min_{U_S} h(A:B)_{\rho_{AB}(U_S)},
\end{equation}
where 
\begin{equation}
    \rho_{AB}(U_S) = \Tr_{C} U_S |\psi\rangle \langle \psi| U^{\dagger}_S.
\end{equation}
In the main text, we have considered $A,B$ being square subregions and $S$ being two circles around the trisections.
Let $ABS$ be the disjoint union of $A,B$ and $S$, then all entanglement quantities can be obtained by the covariance matrix $C_{ABS}$. Under an infinitesimal unitary evolution
\begin{equation}
    U_S = e^{idX_S}, 
\end{equation}
where $dX_S$ is Hermitian, the covariance matrix changes as
\begin{equation}
    dC_{ABS} = i [dX_S, C_{ABS}].
\end{equation}
Note that here we are commuting two matrices with different sizes. In this expression $dX_S$ should be understood as an abbreviation for $dX_S\oplus 0_{AB}$. In the discussion throughout this section, we will abuse the same notation for a smaller matrix $X$ and its extension $X\oplus 0$ to a larger size. 

Now we consider the change of entanglement quantities. Firstly, differentiating Eq.~\eqref{eq:fermionEE} we obtain
\begin{equation}
    dS(A)=\Tr(dC_A h_A),
\end{equation}
where $h_A$ is the entanglement Hamiltonian Eq.~\eqref{eq:h_R}. It follows then
\begin{equation}
    dI(A:B)=\Tr(dC_{AB} h_I),
\end{equation}
where 
\begin{equation}
    h_I=h_A+h_B-h_{AB}.
\end{equation}
Note that again $h_A$ is a shorthand for $h_A\oplus 0_B$ and $h_B$ is a shorthand for $0_A \oplus h_B$. Next we consider the change of $S_R(A:B)$. Differentiating Eq.~\eqref{eq:fermionSR} we obtain
\begin{equation}
\label{eq:dSR}
    dS_R(A:B)=\Tr(dC^{R} h_{AA'}),
\end{equation}
where
\begin{equation}
    h_{AA'} = \log\left(\frac{I_{AA'}-C^R_{AA'}}{C^R_{AA'}}\right).
\end{equation}
and
\begin{equation}
\label{eq:dCR}
        dC^{R}=
    \begin{bmatrix}
    dC_{AB} & d\sqrt{C_{AB}(I-C_{AB})} \\
    d\sqrt{C_{AB}(I-C_{AB})} & -dC_{AB}
    \end{bmatrix}.
\end{equation}
For later convenience, we denote the blocks of $h_{AA'}$ by
\begin{equation}
    h_{AA'}=
    \begin{bmatrix}
    h_{00} & h_{01} \\
    h_{10} & h_{11}
    \end{bmatrix},
\end{equation}
where $h_{00}$ is the submatrix containing rows and columns corresponding to $A$ and $h_{01}$ is the submatrix containing rows corresponding to $A$ and columns corresponding to $A'$, etc. Combining Eqs.~\eqref{eq:dCR} \eqref{eq:dSR} we obtain 
\begin{equation}
\label{eq:dSRblock}
    dS_R(A:B)= \Tr(dC_{AB}(h_{00}-h_{11})+d\sqrt{C_{AB}(I-C_{AB})}(h_{01}+h_{10})),
\end{equation}
The differentiation of $\sqrt{C_{AB}(I-C_{AB})}$ needs special care. Let the eigenvalue equation of $C_{AB}$ be 
\begin{equation}
    C_{AB}|r_{\alpha}\rangle = r_{\alpha} |r_{\alpha}\rangle
\end{equation}
then
\begin{equation}
    C_{AB}(I-C_{AB}) = \sum_{\alpha} q_{\alpha}|r_{\alpha}\rangle\langle r_{\alpha}|
\end{equation}
where
\begin{equation}
    q_{\alpha} = r_{\alpha}(1-r_{\alpha}).
\end{equation}
Let $Q$ denote $C_{AB}(I-C_{AB})$, then
\begin{equation}
\label{eq:dQ}
    dQ = dC_{AB}-C_{AB}dC_{AB}-dC_{AB}C_{AB}.
\end{equation}
Note that
\begin{equation}
    dQ=\sqrt{Q}d\sqrt{Q} +d\sqrt{Q} \sqrt{Q}.
\end{equation}
This implies that
\begin{equation}
d\sqrt{Q}=\sum_{\alpha,\beta} \frac{|r_{\alpha}\rangle\langle r_{\alpha}|dQ|r_{\beta}\rangle\langle r_{\beta}|}{\sqrt{q_{\alpha}}+\sqrt{q_{\beta}}}.
\end{equation}
By Eq.~\eqref{eq:dQ} we have
\begin{equation}
    \langle r_{\alpha}|dQ|r_{\beta}\rangle = (1-r_{\alpha}-r_{\beta})\langle r_{\alpha}|dC_{AB}|r_{\beta}\rangle
\end{equation}
and thus
\begin{equation}
\label{eq:doffdiag}
    d\sqrt{C_{AB}(I-C_{AB})} = \sum_{\alpha\beta} \frac{(1-r_{\alpha}-r_{\beta})|r_{\alpha}\rangle\langle r_{\alpha}|dC_{AB}|r_{\beta}\rangle\langle r_{\beta}|}{\sqrt{r_{\alpha}}+\sqrt{r_{\beta}}}.
\end{equation}
Substituting Eq.~\eqref{eq:doffdiag} into Eq.~\eqref{eq:dSRblock} we obtain
\begin{equation}
    dS_R(A:B)=\Tr(dC_{AB} h_R),
\end{equation}
where
\begin{equation}
    h_R = h_{00}-h_{11}+\sum_{\alpha\beta}\frac{(1-r_{\alpha}-r_{\beta})\langle r_{\beta}|(h_{01}+h_{10})|r_{\alpha}\rangle}{\sqrt{r_{\alpha}}+\sqrt{r_{\beta}}} |r_{\beta} \rangle\langle r_{\alpha}|.
\end{equation}
Finally, the change of $h(A:B) = S_R(A:B) - I(A:B) $ is
\begin{equation}
    dh(A:B)=\Tr(dC_{ABS} (h_R-h_I)) = i\Tr(dX_S[C_{ABS},h_R-h_I]). 
\end{equation}
Therefore the gradient direction is
\begin{equation}
    X_S=-i[C_{ABS},h_R-h_I]_S,
\end{equation}
where the subscript $S$ in the RHS means restricting to the submatrix containing rows and coloumns corresponding to $S$. The disentangler is then
\begin{equation}
    U_S=e^{iX_S\delta t},
\end{equation}
where $\delta t$ can be found by a linesearch which minimizes the objective function $h(A:B)$. Repeating the disentangling until $h(A:B)$ converges, we obtain the optimized $h_S$. 

\section{Free fermion models}
In this section, we review the Hofstadter model and the topological insulator. We then review the solution of the Hofstadter model using standard Fourier transform techniques following Harper and Hofstadter.
\subsection{Hofstader model}
The Hofstader model describes a tight binding model coupled to a perpendicular magnetic field on a square lattice. The Hamiltonian is
\begin{equation}
    H(B,\mu)= -t\sum_{xy} ((c^{\dagger}_{x,y} e^{-iA_x(x,y)} c_{x+1,y} + c^{\dagger}_{x,y} e^{-iA_y(x,y)} c_{x,y+1}) + h.c.)+\mu \sum_{xy} c^{\dagger}_{x,y} c_{x,y}
\end{equation}
where $A(x,y)$ is the vector potential of the magnetic field. In the case of a uniform magnetic field $B$, we may choose the Landau gauge
\begin{equation}
\label{eq:gauge}
    A_x(x,y)=0, ~~ A_y(x,y)=Bx.
\end{equation}
The model exhibits a well-known butterfly structure of the energy levels with respect with the magnetic field. For our purpose, we focus on two simple cases $B=\pi/2$ and $B=\pi/3$, where the chemical potential $\mu$ can be tuned such that a certain number of Chern bands is filled. The ground state is therefore an integer quantum Hall state with a Chern number $C\in \mathbb{Z}$. The boundary theory with the minimal central charge is $|C|$ free compactified bosons, with central charge $c=|C|$. At $B=\pi/2$, there are four bands, and the lowest band has $C=1$. Numerically, the lowest band is filled if we choose $t=1$ and $\mu=2$. At $B=\pi/3$, there are six bands, and the lowest two bands both have $C=1$. 
\subsection{Topological insulator}
A topological insulator can be obtained by stacking two Chern insulators with $C=1$ and $C=-1$,
\begin{equation}
    H_{TI} = H_{\uparrow}(\pi/2, 2)+H_{\downarrow}(-\pi/2,2),
\end{equation}
where the arrows label the layers. There is a time reversal symmetry $T$, where
\begin{equation}
    T c_{\uparrow} T^{-1}= c_{\downarrow},~~~~ T c_{\downarrow}T^{-1}= -c_{\uparrow}
\end{equation}
at each site. Note that $T$ is antiunitary such that $TiT^{-1}=-i$. It can be readily seen that the topological insulator Hamiltonian is time-reversal symmetric, $TH_{TI}T^{-1}=H_{TI}$.

The topological insulator is an example of symmetry protected topological phases. The ground state can be adiabatically connected to the product state if the time reversal symmetry is broken. However, it has a free boson CFT mode with $c_{L}=c_{R}=1$ (and thus $c_{+}=2$ on each of the boundary if the $U(1)$ fermion number conservation and the time reversal symmetry are preserved. One intuitive argument goes as follows. The time reversal operator connects the left and right moving modes on the boundary of the two layers. If the time-reversal symmetry is broken, then it allows for disentangling that gaps out the two boundary modes. Otherwise, the edge modes are robust against local perturbations. 


\subsection{Solving the Hofstadter model}
The Hofstadter model is quadratic in the fermionic variables and can be solved using Fourier transform. Following Harper and Hofstadter, the model behaves very diffrently depending on whether $B$ is rational or irrational. We will only focus on the cases where $B$ is rational. Let $B=p/q$ where $p,q$ are incommensurate integers and choosing the Landau gauge Eq.~\eqref{eq:gauge}, the Hamiltonian is translation invariant in the $y$ direction, and $q$-site translation invariant in the $x$ direction. Therefore we use the Fourier modes
\begin{equation}
    c_{k_x,k_y;n} = \frac{1}{\sqrt{N}}\sum_{x,y} e^{-i((k_x+(n-1)k_0) x+k_y y)} c_{x,y},
\end{equation}
where $k_y\in [0,2\pi),k_x\in [0,2\pi/q)$, $k_0=2\pi/q$ and $n=1,2,\cdots,q$. The operators satisfy the anticommutation relations
\begin{equation}
    \{ c^{\dagger}_{k_x,k_y;n} c_{k'_x,k'_y;n'}\} = \delta_{k_x,k'_x} \delta_{k_y,k'_y} \delta_{n,n'}.
\end{equation}
The Hamiltonian can be expressed as
\begin{equation}
    H = \sum_{n,m}\sum_{k_x,k_y}  c^{\dagger}_{k_x,k_y;n} h_{n,m}(k_x,k_y) c_{k_x,k_y;m},
\end{equation}
where
\begin{equation}
    h_{n,m}(k_x,k_y) = (-2t \cos(k_x+(n-1)k_0) +\mu) \delta_{n,m} -t e^{ik_y}\delta_{n+p,m} -t e^{-ik_y}\delta_{n-p,m},
\end{equation}
where periodic boundary conditions on the indices $n,m$ is assumed (i.e., $q+1$ and $1$ is identified).

One then diagonaize $h_{n,m}(k_x,k_y)$ for each $k_x,k_y$ and obtain the eigenvectors
\begin{equation}
    \sum_{m} h_{n,m}(k_x,k_y) V_{m,l}(k_x,k_y) = \epsilon_l(k_x,k_y) V_{n,l}(k_x,k_y),
\end{equation}
where $1=1,2,\cdots,q$ labels different eigenvectors. One obtains $q$ energy bands $\epsilon_{l}(k_x,k_y)$. The ground state is obtained by filling the states with $\epsilon_{l}(k_x,k_y)<0$. One may tune the chemical potential such that certain bands are filled. 

Finally, we can compute the covariance matrix by
\begin{equation}
    \langle c^{\dagger}_{k_x,k_y;n} c_{k'_x,k'_y;n'} \rangle = \sum_{l}\theta(-\epsilon_{l}(k_x,k_y)) \delta_{k_x,k'_x}\delta_{k_y,k'_y} V^{*}_{nl}(k_x,k_y) V_{n'l}(k_x,k_y)
\end{equation}
and then transforming back to the real space,
\begin{equation}
    \langle c^{\dagger}_{x,y} c_{x',y'}\rangle = \frac{1}{N}\sum_{n,n',l} \sum_{k_x,k_y} e^{-ik_x(x-x')-ik_0((n-1)x-(n'-1)x')-ik_y(y-y')} \theta(-\epsilon_{l}(k_x,k_y)) V^{*}_{nl}(k_x,k_y) V_{n'l}(k_x,k_y)
\end{equation}
One may restrict to $x'=1,2,\cdots q$ and $y'=1$ due to translation invariance. The total numerical cost of computing the covariance matrix is $O(N^2q^2)$. This is much smaller than diagonalization of the full Hamiltonian matrix, which costs $O(N^3)$. In order to compute the covariance matrix of a subsystem, one choose the linear size of the total system minus the linear size of the subsystem to be much larger than the correlation length. Then the finite-size corrections in the total system size is negligible.

\section{Different shapes of disentangler}
\begin{figure}
    \centering
    \includegraphics[width=0.5\linewidth]{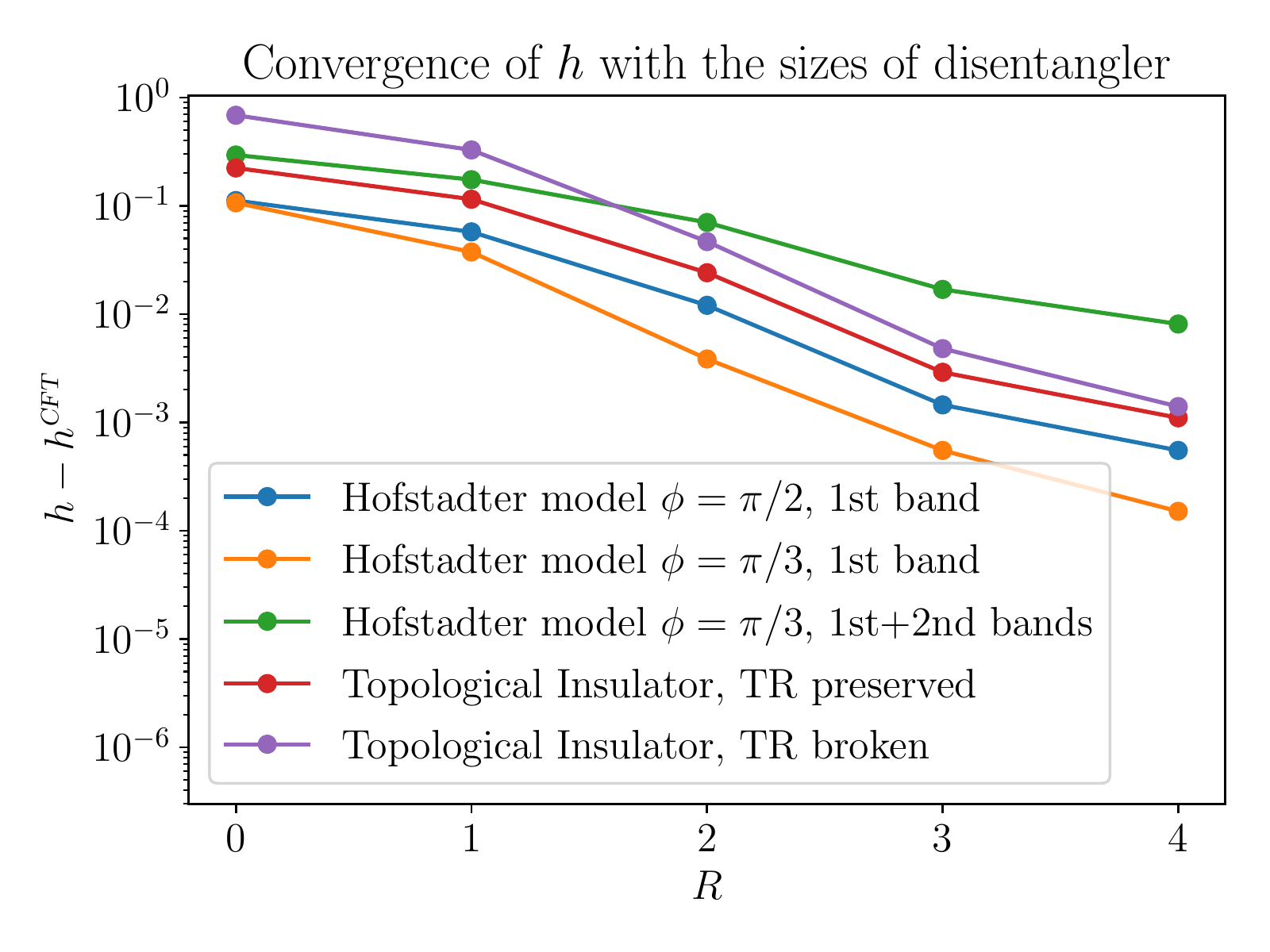}
    \caption{Convergence of $h$ with respect to the disentangler size $R$ for different models using the $U_N U_S$ disentangler.}
    \label{fig:hconv}
\end{figure}

In this appendix we show the numerical value of
\begin{equation}
    h = \min_{U} h(A:B)_{U|\psi\rangle_{ABC}},
\end{equation}
where $|\psi\rangle$ is the ground state of a 2D lattice model, and $A,B$ are adjacent square regions in the bulk, as defined in the main text. We first focus on the case where $U = U_N U_S$, where $U_N,U_S$ are unitaries supported on the circles near the trisections. We vary the radius $R$ of the circles and observe that $h$ converges to $h^{\CFT}\equiv\frac{c_{+}}{3}\log 2$ exponentially with $R$. The numerical result for different models and different $R$'s is listed in Tab.~\ref{tab:radius} and the finite-size corrections are shown in Fig,~\ref{fig:hconv}

Next, we consider alternative forms of the disentangler. In particular, we consider $U=U_{NS}$ that is supported on the union of the circles around the two trisections. In this case the disentangler allows further disentangling between the left and right moving modes and we expect that $h=0$ regardless of whether there are chiral edge modes. Indeed we find numerically that $h \approx 0$ for all cases considered. We also consider the strip disentangler, $U=U_{strip}$, where $U_{strip}$ is supported on a strip with length $L_A$ and width $2R$ that is symmetric along the line connecting $N$ and $S$. The reason to consider the strip disentangler is to see whether the universal universal tripartite entanglement $h_{\mathrm{IR}}$ can be obtained by $\rho_{AB}$ alone (Recall that $U_{N}U_S$ has support inside $C$). As it turns out, even though the strip disentangler gives $h=h^{\CFT}$ for some models considered here, it does not work for a generic model. For example, for the topological insulator with TR broken, we have $c=0$ on the boundary but still we obtain significant $h$ using the strip disentangler.  The result is summarized in Tab.~\ref{tab:geometries}.
\begin{table}[]
    \centering
    \begin{tabular}{|c|c|c|c|c|c|c|c|}
    \hline
        system & $c_{+}$ & $\frac{c_{+}}{3}\log 2 $ & $h_{R=0}$ & $h_{R=1}$ & $h_{R=2}$ & $h_{R=3}$ & $h_{R=4}$ \\ \hline
        Hofstadter model $B=\pi/2$, lowest band & 1 & 0.2310 &0.3429 & 0.2885&0.2431 & 0.2325  & 0.2316
        \\ \hline
        Hofstadter model $B=\pi/3$, lowest band & 1 & 0.2310 & 0.3377& 0.2685 & 0.2349& 0.2316&0.2312
        \\ \hline
        Hofstadter model $B=\pi/3$, lowest two bands & 2 & 0.4621 & 0.7567&0.6365 & 0.5323&0.4790 &0.4702
        \\ \hline
        Topological insulator (TR preserved) & 2 & 0.4621 & 0.6857 & 0.5771 & 0.4862 & 0.4650 & 0.4632
        \\ \hline
        Topological insulator (TR broken) & 0 & 0 & 0.6857&0.3280 & 0.0467 & 0.0048 & 0.0014
        \\ \hline
    \end{tabular}
    \caption{$h$ from disentanglers $U_N,U_S$ supported on different radius. Note that at most $R=4$ is shown here, as opposed to the main text where we used $R=6$ for the Hofstadter model with $B=\pi/3$ and lowest two bands filled. Tolerance of the norm of graident is $\eta<3\times 10^{-3}$.}
    \label{tab:radius}
\end{table}

\begin{table}[h!]
    \centering
    \begin{tabular}{|c|c|c|c|c|c|}
    \hline
        system & $c_{+}$ & $\frac{c_{+}}{3}\log 2 $ & $h$ from $U_N U_S$ & $h$ from $U_{NS}$ & $h$ from $U_{strip}$    \\ \hline
        Hofstadter model $B=\pi/2$, lowest band & 1 & 0.2310 & 0.2316& 0.0007 & 0.2330
        \\ \hline
        Hofstadter model $B=\pi/3$, lowest band & 1 & 0.2310 & 0.2312& 0.0007  & 0.2321
        \\ \hline
        Hofstadter model $B=\pi/3$, lowest two bands & 2 & 0.4621 & 0.4702 & 0.0007 & 0.4829
        \\ \hline
        Topological insulator (TR preserved) & 2 & 0.4621 & 0.4632 & 0.0014 & 0.4660
        \\ \hline
        Topological insulator (TR broken) & 0 & 0 & 0.0014 & 0.0014 & 0.4656
        \\ \hline
    \end{tabular}
    \caption{$h$ from different shapes of disentanglers. The circles on which $U_N,U_S$ are supported have radius $R=4$ and the strip has width $2R=8$.}
    \label{tab:geometries}
\end{table}

\begin{figure}[h!]
    \centering
    \includegraphics[width=0.49\linewidth]{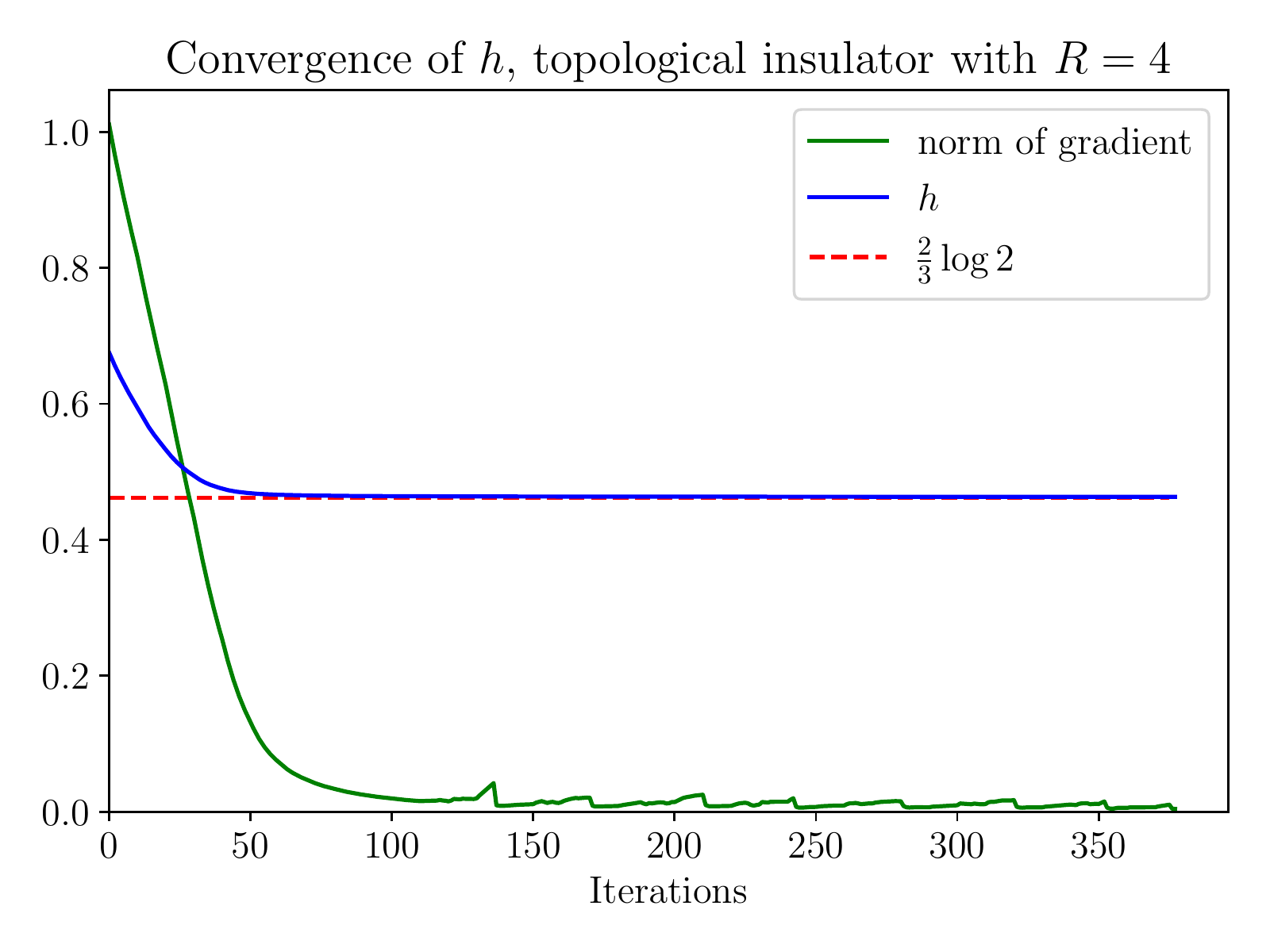}
    \includegraphics[width=0.49\linewidth]{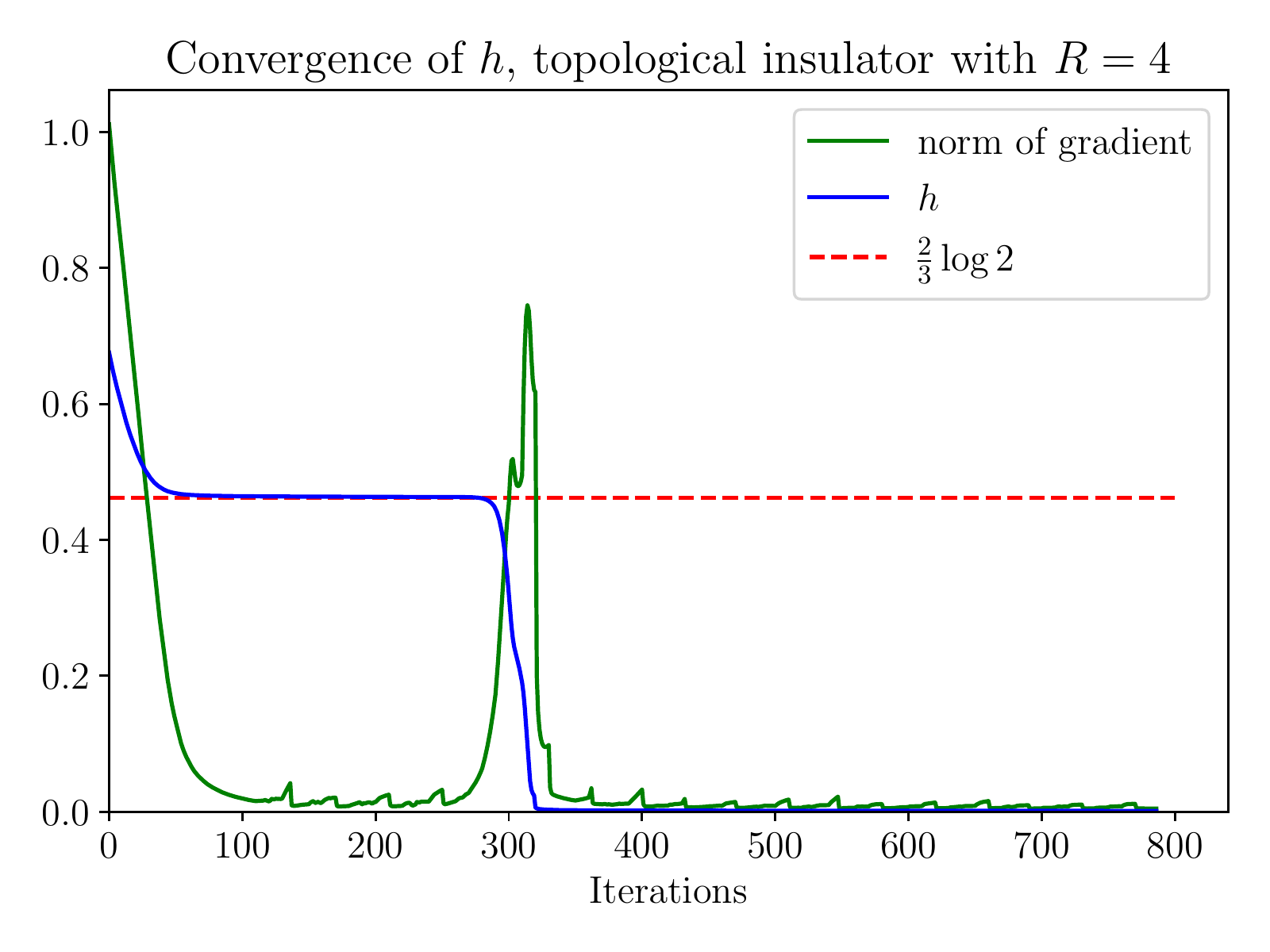}
    \caption{Convergence of $h(A:B)$ in the optimization of disentanglers for the topological insulator. Left: using TR preserved unitaries $U_N U_S$ with circle radius $R=4$. Right: using TR broken unitaries $U_N U_S$ with circle radius $R=4$.}
    \label{fig:conv}
\end{figure}

Finally, we note that in the optimization of $h$ for the topological insulator with TR broken, we add a bit randomness to the gradient descent algorithm to break TR and avoid saddle points. It is expected that there is a saddle point at $h=\frac{2}{3}\log 2$ which corresponds to the TR preserved minimum. Indeed, in an actual simulation we find that the optimization gets stuck at $h\approx \frac{2}{3}\log 2$ for some time with a small norm of gradient but finally reaches $h\approx 0$ after getting around the saddle point (see Fig.~\ref{fig:conv}).

\section{Review of Formalism of String-Net Liquids}
We briefly review the construction of string net liquids. String-net models are exactly solvable lattice models, whose ground state wavefunctions can be thought of as fixed-point wavefunctions of topological phases of matter~\cite{Levin2005}.
They take in a set of diagrammatic rules (formally a spherical pivotal fusion category) as input and results in a topological quantum field theory (TQFT) [formally a unitary modular tensor category (UMTC)], i.e.\ a topological phase.
They classify all gappable topological orders in 2D~\cite{freed2021gapped}.

For simplicity, in the review below, we concentrate on a subset of topological orders without fusion multiplicities, and with vertex operator basis satisfying strict rotation and isotopy invariance.  More general topological orders can be understood through the formalism of fusion categories, see Refs.~\onlinecite{Kitaev2006,Bonderson:Thesis:07,Bonderson09,kassel2012quantum} for an introduction.

\begin{figure}
    \centering
    \includegraphics[width=0.4\linewidth]{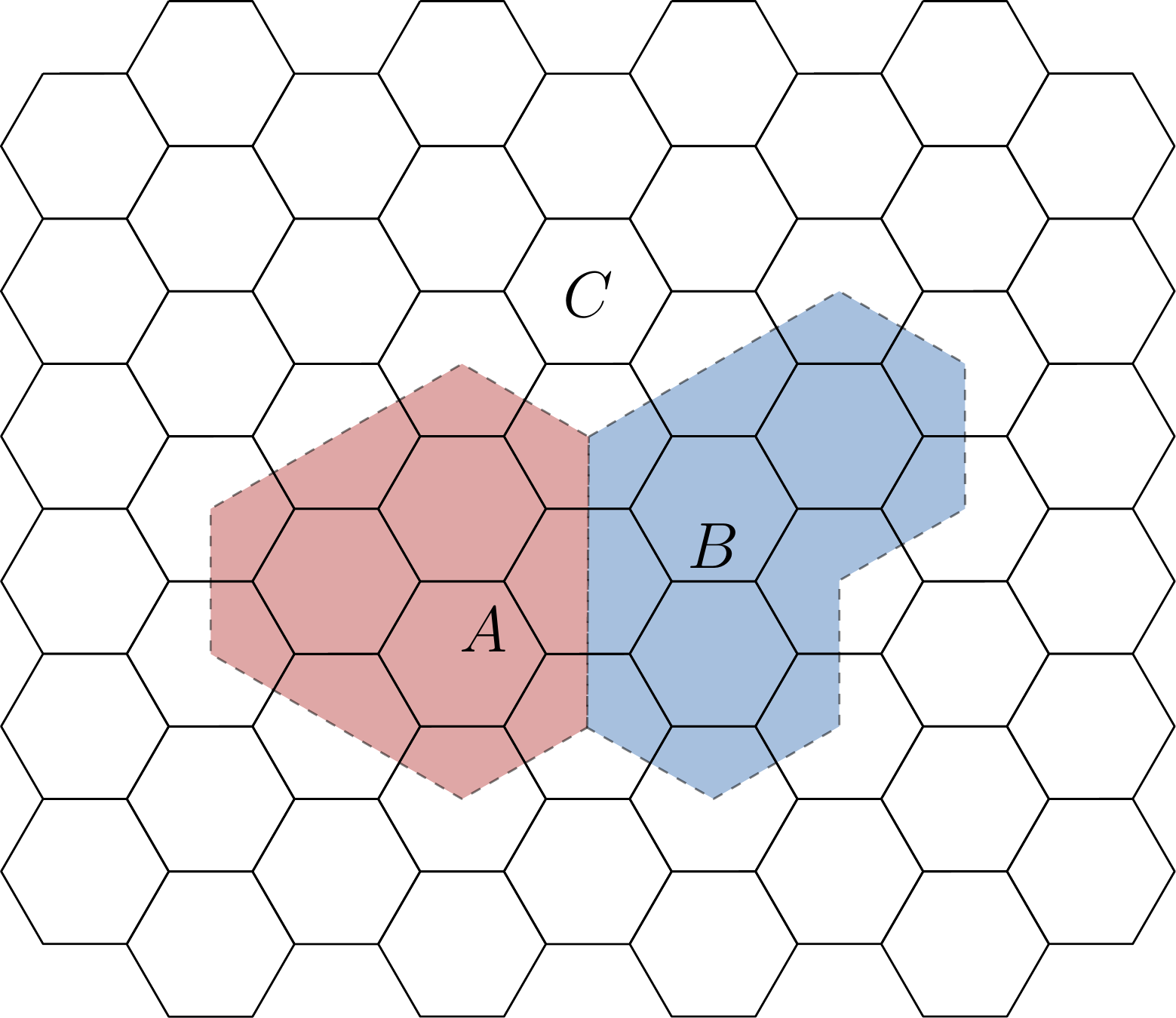}
    \caption{Example tripartition of the string-net liquid on a honeycomb lattice}
    \label{fig:string_net_tripartition}
\end{figure}

A 2D string-net wavefunction is defined on a trivalent lattice where degrees of freedom live on the edges of the lattice such as in  Fig.~\ref{fig:string_net_tripartition}. The local Hilbert space on each edge is spanned by $N+1$ basis states $i={0, 1, ..., N}$ that correspond to ``string types'', each associated with a real number $d_i$ such that $|d_i|>0$ known as that string type's quantum dimension and define $\alpha_i = \sgn(d_i)$.

The strings are allowed to ``branch'' according to ``branching rules'' or ``fusion constraints'' $\delta_{ijk}$ which is 1 if string types $ijk$ can meet at a vertex and 0 otherwise. The ground state wavefunction is characterized by a set of graphical rules relating different string configurations. A central object in that graphical rule is a six-index tensor called the $F$-symbol, denoted $F^{ijm}_{kln}$, which satisfies certain compatibility requirements. Given $F$-symbols and string types, we can construct an exactly solvable projector Hamiltonian, the ground states of which satisfy the graphical rules. 

We briefly summarize some properties of the string net wavefunctions and properties of isotopy-invariant $F$-symbols. The quantum dimension of each string $s$ is  encoded in the $F$-symbol as $1/F^{s\bar{s}0}_{s\bar{s}0}$ and satisfies the following equation:
\begin{equation}
    \sum_k \delta_{ij\bar{k}} d_k = d_i d_j
    \label{eq:sum_quantum_dims}
\end{equation}
$F$-symbols also encode the fusion constraints as follows:
\begin{equation}
    F^{ijm}_{kln} = F^{ijm}_{kln} \delta_{ijm} \delta_{kl\bar{m}}\delta_{iln}\delta_{jk\bar{n}}
    \label{eq:F_symbol_fusion}
\end{equation}
They satisfy the \emph{pentagon equation}
\begin{equation}
    \sum_{n=0}^N F^{mlq}_{k\bar{p}n} F^{jip}_{mn\bar{s}} F^{j\bar{s}n}_{lk\bar{r}} = F^{jip}_{\bar{q}k\bar{r}} F^{ri\bar{q}}_{ml\bar{s}} .
    \label{eq:pentagon_equation}
\end{equation}
Defining $v_i = \sqrt{|d_i|}$, they can be normalized as 
\begin{equation}
    F^{ijk}_{\bar{j}\bar{i}0} = \frac{v_k}{v_i v_j} . \delta_{ijk}
    \label{eq:F_symbol_normalization}
\end{equation}
From here, it can be shown that
\begin{equation}
    \alpha_i \alpha_j \alpha_k = 1
    \label{eq:FS_fusion_constraint}
\end{equation}
if $\delta_{ijk} = 1$. They obey so-called tetrahedral symmetry:
\begin{equation}
    F^{ijm}_{kln} = F^{jim}_{lk\bar{n}} = F^{lk\bar{m}}_{jin} = \frac{v_m v_n}{v_jv_l} F^{imj}_{\bar{k}nl}
    \label{eq:tetrahedral_symmetry}
\end{equation}
Finally, the $F$-symbols must satisfy a ``unitarity" constraint:
\begin{align}
    (F^{ijm}_{kln})^* = F^{\bar{i}\bar{j}\bar{m}}_{\bar{k}\bar{l}\bar{n}}
    \label{eq:F_symbol_unitarity}
\end{align}
which yields, by Eq.~\ref{eq:pentagon_equation}, 
\begin{equation}
    \sum_{n=0}^N (F^{ijm'}_{kln})^*F^{ijm}_{kln} = \delta_{m,m'} \delta_{ijm} \delta_{kl\bar{m}}
    \label{eq:F_symbol_orthogonality}
\end{equation}
\begin{figure}
    \centering
    \includegraphics[width=0.4\linewidth]{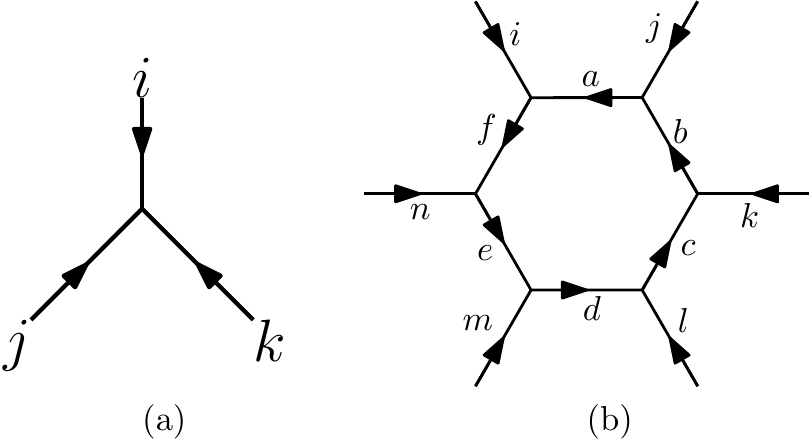}
    \caption{Hamiltonian operators (a) Vertex for $A_v$ terms, which enforce fusion contraints (b) Plaquette for $B_p$ terms}
    \label{fig:Hamiltonian_operators}
\end{figure}
The string-net liquids are ground states of commuting projector Hamiltonians which may be written in terms of fusion contraints and $F$-symbols. It is a sum of fusion constraints associated to each vertex as shown in Fig.~\ref{fig:Hamiltonian_operators} (a):
\begin{equation}
    A = \sum_{v}\delta_{ijk}
    \label{eq:vertex_operators}
\end{equation}
and plaquette operators which introduce ``dynamics" as in Fig.~\ref{fig:Hamiltonian_operators} (b):
\begin{align}
    B &= \sum_{p}B_p \\
    B_p &= \sum_{s} \frac{d_s}{D}B_p^s \\
    B_p^s &= \prod_{\{i, a,b\}\in v\in p} F^{i\bar{a}b}_{sb'\bar{a}'} \delta_{i,i'}
    \label{eq:plaquette_operators}
\end{align}
The total Hamiltonian is then 
\begin{equation}
    H = A + B
\end{equation}
We will consider a sequence of transformation which will modify the underlying triangulation and therefore modify the Hilbert space on which the Hamiltonian is defined, so we denote a triangulation by $\Gamma$ and the Hamiltonian defined on it as $H(\Gamma)$.

\section{$h(A:B)$ for gappable topological order}

In this Appendix, we show that $h(A:B)=0$ for a tripartition of a wavefunction with gappable topological order using the string-net liquid formalism. We first review the definitions of triangle states and sums of triangle states (SOTS) from Ref.~\onlinecite{Zou2021}.

\begin{definition}[Triangle state]\label{def:triangle_state} A pure tripartite state $\ket{\psi}_{ABC} \in \calH_A\otimes\calH_B\otimes\calH_C $ is a triangle state if for each local Hilbert space there exists a bipartition $\calH_\alpha =(\calH_{\alpha_L}\otimes \calH_{\alpha_R}) \oplus \calH_{\alpha}^0$ ($\alpha =A,B,C$) such that
\begin{equation}
	\ket{\psi}_{ABC}= \ket{\psi}_{A_RB_L}\ket{\psi}_{B_RC_L}\ket{\psi}_{C_RA_L} .
	\label{eq:triangle_app}
\end{equation}
\end{definition}
\begin{definition}[Splitting]\label{def:splitting}
A splitting of a Hilbert space $\calH_i$ is an orthogonal decomposition of the Hilbert space into a direct sum of tensor product spaces 
\begin{align}
    \calH_i = \calH_i^0 \;\oplus\; \bigoplus_j \calH_{iL}^j \otimes \calH_{iR}^j \,.
\end{align}
\end{definition}
\noindent The space $\calH_i^0$ may be 0-dimensional.

\begin{definition}[Sum of polygon states---SOPS]\label{def:SOPS}
An $N$-party pure quantum state $\ket{\psi} \in \calH_1 \otimes \calH_2 \otimes \cdots \otimes \calH_N$ is a SOPS with respect to the decomposition $(\calH_1,\calH_2,\dots,\calH_N)$ if for each party $i$, $\calH_i$ admits a splitting and
\begin{align}
	\ket{\psi} &= \sum_j a_j \bigotimes_i \ket{j}_{(iR)(i^+L)}
	\qquad \mathrm{such\ that\ } \ket{j}_{(iR)(i^+L)} \in \calH_{iR}^j \otimes \calH_{i^+L}^j \,.
	\label{eq:SOPS_definition}
\end{align}
where $i^+ \equiv (i\bmod N) +1$ denote the party after $i$, the coefficients are normalized to $\sum_j |a_j|^2 = 1$.
\end{definition}

An SOTS is then an SOPS for $N=3$. It was shown in Ref.~\cite{Zou2021} that a pure tripartite quantum state is a sum of triangle states as defined in Def.~\ref{def:SOPS} if and only if $h(A:B) = 0$. To show that $h(A:B)=0$ for string-net liquids, it suffices then to show that the fixed point wavefunctions are SOTS. 

We consider a tripartition of a string-net liquid on a trivalent lattice covering the surface of a sphere, shown schematically in Fig.~\ref{fig:string_net_tripartition}. We pursue an approach similar to Ref.~\cite{Levin2006} in which we start with the string-net liquid on a trivalent lattice covering a sphere, reduce the interior of each subregion to a tree-like lattice connecting it to its neighbors, and finally analyze the wavefunction of this state.

The multipartite entanglement measure $h(A:B)$ of a  tripartite quantum state is invariant under local isometries, however acting jointly on two out of three parties can change its value. We must be mindful that all operations remain local to each region $A$, $B$, and $C$, just as the reduction in Ref.~\onlinecite{Levin2006} is restricted to the regions so that the state can only be reduced to tree-like diagrams rather than the vacuum altogether. 

To perform this reduction formally, we need to ensure that there is a unitary operation which can disentangle degrees of freedom from within each region $A$, $B$, and $C$. This can be achieved using the graphical operations in Eqs. (4) through (7) of Ref.~\cite{Levin2005} if we can promote the so-called $F$-move (in Eq. (7)), denoted below by $\bfP$, to a unitary operation, denoted below by $\tilde{P}$, as the remaining operations derive from it. We first explicitly state the matrix elements of $\bfP$:

\begin{equation}
    \left\langle\vcenter{\hbox{\includegraphics[scale=0.4]{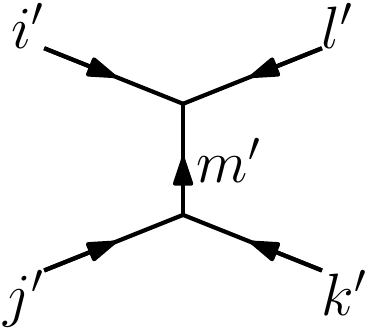}}} \right\rvert\bfP\left\rvert\vcenter{\hbox{\includegraphics[scale=0.4]{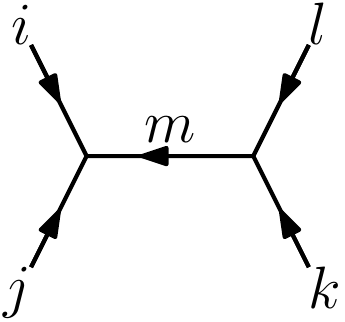}}}\right\rangle \equiv F^{ijm}_{klm'} \delta_{i,i'}\delta_{j,j'}\delta_{k,k'}\delta_{l,l'}
\end{equation}

Let $\bbV$ denote the Hilbert space of the five links on which $\bfP$ acts and let $\bbV = \bbV_f \oplus \bbV_0$ denote an orthogonal decomposition of $\bbV$ into the space of configurations which satisfy the fusion constraints and the space of configurations which do not. Then because the $F$-symbol encodes fusion constraints, $\ker(\bfP) = \bbV_0$. Using Eq.~\ref{eq:F_symbol_orthogonality}, we can show that 
\begin{equation}
    \bfP^\dagger \bfP  = \delta_{m,m''} \delta_{ijm}\delta_{kl\bar{m}}
    \label{eq:almost_isometry_P}
\end{equation}
In other words, it is a projector on to $\bbV_f$. We can promote it to a unitary on $\bbV$ by extending it to a new operation $\tilde{\bfP}$ as follows:
\begin{equation}
    \tilde{\bfP} \equiv \bfP \oplus U
\end{equation}
where $U$ is an arbitrary unitary on $\bbV_0$.

\begin{figure}
    \centering
    \includegraphics[width=0.5\linewidth]{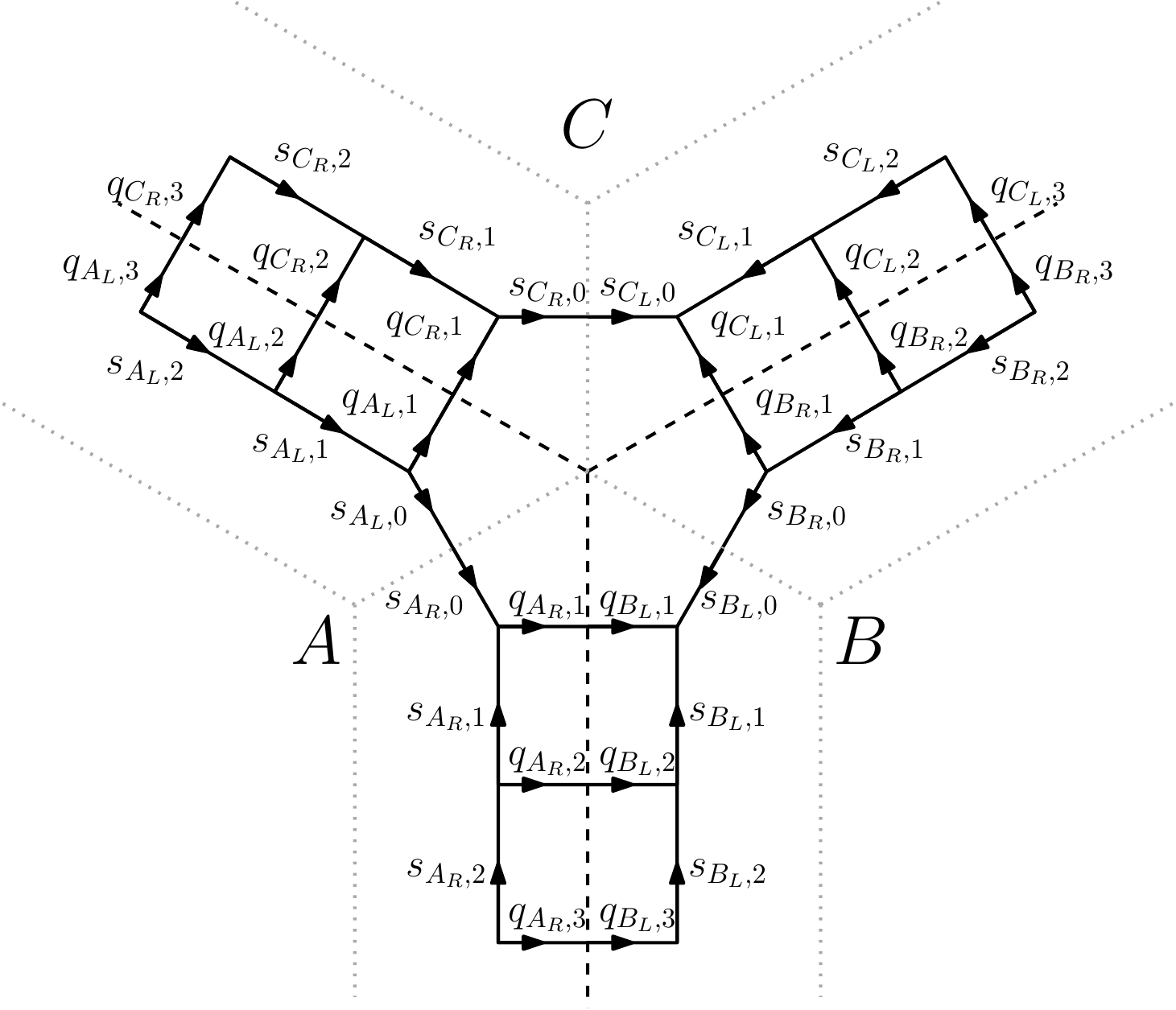}
    \caption{Simplified string-net liquid after application of local isometries on regions $A$, $B$, and $C$. A 0-string (not shown) connects the outer most vertices of each region, e.g. the vertex with $q_{A_L,3}$ and $s_{A_L,2}$ coming out also has a 0-string connecting it to the vertex with $q_{A_R,3}$ and $s_{A_R,2}$ coming out.}
    \label{fig:simplified_tripartite_SNL}
\end{figure}

By repeatedly performing $\tilde{\bfP}$ (e.g. until one has a bubble connected to two legs, then removing the bubble), it is possible to disentangle degrees of freedom from the interior of each region $A$, $B$, and $C$. As in Ref.~\cite{Levin2006}, we are left with tree-like diagrams on the boundaries of each region and the reduced global wavefunction may be represented as in Fig.~\ref{fig:simplified_tripartite_SNL}

We now study the reduced wavefunction and show that it is of SOTS form. We can again employ the graphical identities in Eqs.~(4)-(7) in Ref.~\cite{Levin2005} to spot some properties of this wavefunction as in Ref.~\onlinecite{Levin2006}. 

First, consider the local bipartitions as shown by the dotted gray lines in Fig.~\ref{fig:simplified_tripartite_SNL}, e.g.\ $A = A_L \otimes A_R$. We can see immediately that for any two local bipartitions which straddle the tripartition (shown by the solid dashed line), such as $A_L$ and $C_R$, the $q$ degrees of freedom for any two adjacent local bipartition must be equal, e.g. $\vec{q}_{A_L}$ must equal $\vec{q}_{C_R}$. 

Next, by applying the $F$-move, we can see that $s$ degrees of freedom between two bipartitions which are adjacent across the tripartition must also be equal, e.g. $\vec{s}_{A_L}$ must equal $\vec{s}_{C_R}$. Finally, we can see that all of the $s$ degrees of freedom on the center hexagon must be equal. This is the origin of the SOTS structure: the $s$ degrees of freedom form an effective GHZ-like state, and the remaining degrees of freedom dress it according to fusion constraints. 
\begin{equation}
    \ket{\psi(s)}_{A_LC_R} \equiv \ket{s}_{A_{L,0}}\ket{s}_{C_{R,0}}\ket{\psi(s)}_{A_{L,1:n}C_{R,1:n}}
\end{equation}    
where $\ket{\psi(s)}_{A_{L,1:n}C_{R,1:n}}$ is, up to normalization,
\begin{equation}
    \ket{\psi(s)}_{A_{L,1:n}C_{R,1:n}} = \sum_{\substack{\vec{q}_{A_{L,1:n}}\\\vec{q}_{C_{R,1:n}}\\\vec{s}_{A_{L,1:n}}\\\vec{s}_{C_{R,1:n}}}} \delta_{\vec{q}_{A_{L,1:n}}\vec{q}_{C_{R,1:n}}}\delta_{\vec{s}_{A_{L,1:n}}\vec{s}_{C_{R,1:n}}}\left(\prod_{i=1}^{N-1}\sqrt{|d_{q_i}|}\right)\ket{\vec{q}}_{A_{L,1:n}}\ket{\vec{q}}_{C_{R,1:n}}\ket{\vec{s}}_{A_{L,1:n}}\ket{\vec{s}}_{C_{R,1:n}} .
\end{equation}
The remaining two wavefunctions, $\ket{\psi(s)}_{B_LA_R}$ and $\ket{\psi(s)}_{C_LB_R}$ are defined similarly through the index $s$. The total wavefunction is then, up to normalization,
\begin{equation}
    \ket{\psi} = \sum_s d_s \ket{\psi(s)}_{A_LC_R}\ket{\psi(s)}_{B_LA_R}\ket{\psi(s)}_{C_LB_R}
\end{equation}
which satisfies Def.~\ref{def:SOPS}. We therefore conclude that $h(A:B)=0$ for string-net liquids.

\end{document}